\documentstyle[11pt,aaspp4]{article}

\newcommand\msun{$M_{\odot}\  $}

\slugcomment{Accepted for publication in ApJ.}

\lefthead{Wen  et al.}
\righthead{ X-ray Orbital Modulation of Cygnus X-1} 

\begin{document}

\begin{center}
\title{Orbital Modulation of X-rays from Cygnus X-1 in its Hard and Soft States}
\author{Linqing Wen \altaffilmark{1}, Wei Cui, Alan M. Levine, and  Hale V. Bradt}
\affil{Center for Space Research, MIT, Cambridge, MA 02139 USA}
\altaffiltext{1}{email lqw@space.mit.edu}


\end{center}

\begin{abstract}

We have analyzed over $2$ years of $\it RXTE$/ASM data for Cygnus
X-1. We have detected the $5.6$-day orbital period in Lomb-Scargle periodograms of both light curves and hardness ratios
when Cyg X-1 was in the hard state. This detection was made with
improved sensitivity and temporal coverage compared with previous
detections by other X-ray missions. The folded light curves and
hardness ratios show a broad intensity dip accompanied by spectral
hardening centered on superior conjunction of the X-ray source. The
dip has a duration of about 27\% of the orbital period and depth
ranging from $8$\% to $23$\% of the non-dip intensities in three
energy bands. Variability on time scales of hours is often evident
within the broad dip in the unfolded data. In contrast, no feature at
the orbital period is evident in the periodograms or folded
light curves for the soft state. Absorption of X-rays by a stellar
wind from the companion star can reproduce the observed X-ray orbital
modulations in the hard state.  To explain the low orbital
modulation in the soft-state data, a reduction of the wind
density during the soft state would  be required.  As an alternative,
a partial covering scenario is described which could also account
for the lack of the orbital modulation in the soft state.

\end{abstract}


\section{Introduction}

Cyg X-1 has been identified as a binary system of $5.6$-day orbital period
which contains an O$9.7$ Iab supergiant and a compact object that is
believed to be a black hole (\cite{bolton72}; \cite{webster72}). The
observed intense X-ray flux from this system is thought to be produced
close to the black hole in an accretion disk which emits soft X-ray photons
and in a hot corona ($T\sim 10^8$--$10^9$ K) that inverse-Compton
scatters low energy photons to higher energies (e.g., \cite {nolan84};
\cite{tanaka95} and references therein). The accretion flow from the
supergiant is probably intermediate between
Roche-lobe overflow and stellar wind accretion (e.g., \cite{gies86b}).

\begin{figure}[f]
\begin{centering}
\epsfxsize=6.5in \epsfbox{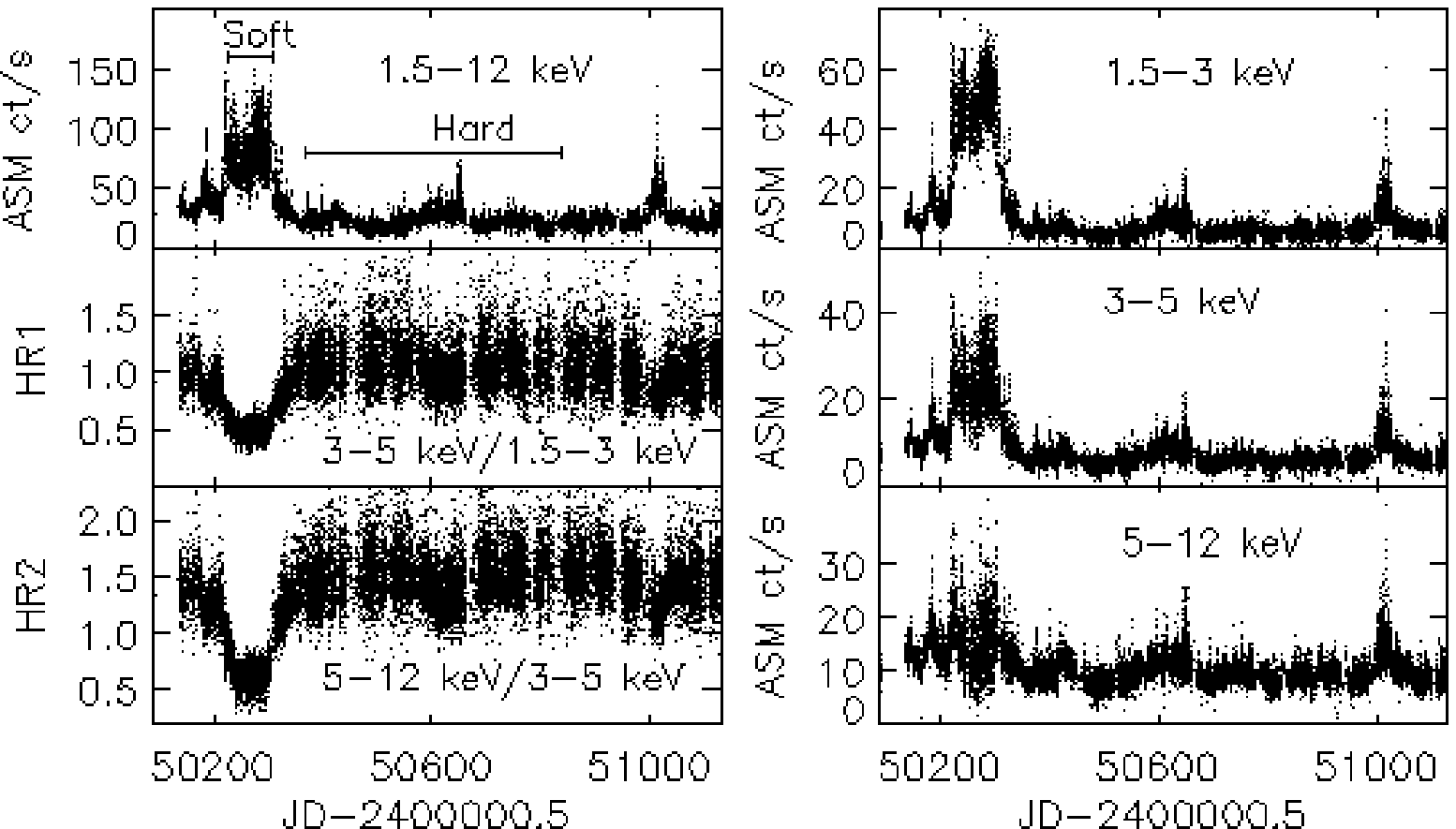}
\end{centering}
\figcaption[fig1.eps]{$\it RXTE$/ASM light curves and hardness ratios of Cyg X-1 for the
period $1996$ March to $1998$ August.  MJD $51000$ corresponds to
$1998$ July $6$. The marked intervals indicate the hard-state  and soft-state data used for the analysis.  An $80$-day soft state is apparent. \label{ltcall}}
\end{figure}

Two physically distinct states of Cyg X-1 have been observed: the hard
state and the soft state.  Most of the time, Cyg X-1 stays in the hard
state where its $2$--$10$ keV luminosity is low and the energy
spectrum is hard. Every few years, Cyg X-1 undergoes a transition to
the soft state and stays there for weeks to months before returning
to the hard state.  During the transition to the soft state, the
$2$--$10$ keV luminosity increases, often by a factor of more than
$4$, and the energy spectrum becomes softer (see reviews by
\cite{oda77}; \cite{nolan84} and references therein; also the $\it
RXTE$/ASM light curves in Fig. \ref{ltcall}). Interestingly, the total
$1.3$--$200$ keV luminosity remained unchanged to within $\sim 15\%$
during the $1996$ hard-to-soft and soft-to-hard state transitions (\cite{zhang97}).

The hard state of Cyg X-1 frequently exhibits short, irregular,
absorption-like X-ray intensity dips.  These dips usually last
seconds to hours and seem to occur preferentially near superior
conjunction of the X-ray source. They are often thought to be due to
absorption in inhomogeneities in the stellar wind from the companion
(e.g., \cite{pravdo80}; \cite{ron84}; \cite{kitamoto84};
\cite{balucinska91}; \cite{ebisawa96}). 

 The most probable mass for the black hole is about $10$ \msun  (\cite{herrero95}; see also \cite{gies86a} for a slightly higher value).   One of the larger
uncertainties in the determination of the mass comes through the inclination angle
$i$, which remains relatively poorly constrained.  The various  existing
 techniques to determine $i$, such as those using the variation
of the polarization of the optical light, allow it to be in a wide
range of $25$--$70^{\circ}$ (e.g., \cite{long80}). 

The $5.6$-day orbital period of Cyg X-1 may be detected via several effects.  In the
optical band, this period manifests itself as radial velocity
variations of the absorption/emission lines (\cite{bolton75}) and as
ellipsoidal light variation (e.g., \cite{walker72}).  Phase-dependent
variations of the equivalent width of the UV lines of Si IV and C IV
have been reported by Treves (1980) and attributed to the orbital
motion of the X-ray heated region of the stellar wind. Orbital
modulations in the near-infrared $J$ and $K$ band (\cite{leahy92}) and
in radio at $15$ GHz (\cite{pooley99}) have also been reported.  The
causes of these modulations are still speculative.

X-ray orbital modulations in the hard-state data of several
investigations show an intensity minimum around superior conjunction
in the folded light curves.  A $1300$-day record of Ariel 5 ASM
observations in the $3$--$6$ keV band yielded  an
intensity minimum near superior conjunction even though the $5.6$ day
period was not detected at a convincing level of statistical
significance in a power density spectrum (\cite{holt79}).  The existence of a broad
dip near superior conjunction was confirmed in $100$ days of hard
state data from the WATCH/Eureca wide field X-ray monitor
(\cite{priedhorsky95}). In the $9$--$12$, $12$--$17$ and $17$--$33$
keV bands, the dips had depths of $\sim 21\%$,  $20\%$, and $10\%$
respectively.  The width (FWHM) is $26\%$ of the period in the
$9$--$12$ keV band.  It is unclear how this type of broad dip is
related to the shorter irregular dips discussed previously.  A $5\%$
peak-to-peak orbital modulation was also found in 3 years of BATSE
data in the $45$--$200$ keV band (\cite{robinson96}).

In this paper, we present a detailed study of the orbital modulation
in the $1.5$--$12$ keV  energy band using data from the All-Sky
Monitor (ASM) on board the {\it Rossi X-ray Timing Explorer} ($\it
RXTE$).  The
advantages of $\it RXTE$/ASM observations lie in its relatively good sensitivity ($\le 10$ mCrab in a day), frequent data sampling ($\sim 10$--$20$ times a day) and long baseline ($\ge 2.5$ years).
Moreover, the fact that a long ($\sim 80$ days) soft state was observed
by the ASM in $1996$ makes it possible to quantitatively compare the
two states.  An earlier report of the detection of the $5.6$ day
period in the $\it RXTE$/ASM data was made by Zhang et al. (1996).

 Our analysis focuses on the X-ray orbital modulation with
the goal of investigating the cause of the broad intensity dip,  an understanding of which may ultimately help constrain  the
system parameters.  Specifically, we present (1) the results of a periodicity search; (2) the folded and individual orbital light curves; (3) a comparison of
the orbital modulations in the soft and hard states; and (4) the
results from a simulation of the orbital modulation caused by a 
partially ionized stellar wind from the companion.

\section{Data}

The All-Sky Monitor on board $\it RXTE$ (\cite {bradt93}) has been
monitoring the sky routinely since $1996$ March.  The ASM consists of
three Scanning Shadow Cameras, each consisting of a coded mask and a
position-sensitive proportional counter. A linear least squares fit to
the shadow patterns from a $90$-s observation by one of the three
cameras of the ASM yields the source intensity in three energy bands
($1.5$--$3$, $3$--$5$, and $5$--$12$ keV).  The intensity is usually
given in units of the count rate expected if the source were at the
center of the field of view in one of the cameras; in these units, the
$1.5$--$12$ keV Crab nebula flux is about $75$ ASM ct/s. The estimated
errors of the source intensities include the uncertainties due to
counting statistics and a systematic error taken to be $1.9\%$ of the
intensities.  A source is typically observed $10$--$20$  times a
day.  In the present analysis, we have used source intensities of 90-s time resolution derived at MIT by the {\it RXTE}/ASM  team. A detailed description of the ASM and
the light curves can be found in Levine et al. (1996) and Levine (1998).

The X-ray light curves and hardness ratios from the ASM observations
of Cyg X-1 ($1996$ March -- $1998$ September) are shown in
Fig. \ref{ltcall}.  During $1996$ May ( MJD $\sim 50220$, where MJD=JD-2400000.5 ), a transition into the soft
state is evident (\cite{cui96}). After about $80$ days in the soft state, Cyg X-1 returned to the hard state and remained there through 1998 September.
The hard-state light curve shows long-term variations on time scale of
$100$--$200$ days and rapid flares that seem to occur every $20$ to
$40$ days (see also \cite{cui98}).  For the analyses discussed below,
the hard-state data are taken from a $470$-day interval (MJD
$50367.432$--$50837.324$) and the soft-state data from a $80$-day
interval (MJD $50227.324$--$50307.324$).

In this paper, the hardness ratio HR$1$ is defined as the
ratio of the ASM count rates in the $3$--$5$ keV band to that of the
$1.5$--$3$ keV band,
and the hardness ratio HR$2$ as the ratio of the count rates of the
$5$--$12$ keV band to that of the $3$--$5$ keV band.

\section{Analysis and Results}

Periodicities in both  the hard state and the soft state have been sought by means of Lomb-Scargle periodograms of both the light curves and the derived
hardness ratios.  The Lomb-Scargle periodogram (\cite{lomb76};
\cite{scargle82}; \cite{press92}) was used to estimate the power density spectrum instead of the classic
periodogram based on the Fast Fourier Transform (FFT) 
since the ASM data points are unevenly spaced in time.  In the Lomb-Scargle periodogram, a maximum in the power occurs at the frequency which
gives the least squares fit of a sinusoidal wave to the data. We
oversampled the spectrum so that the frequencies are more closely
spaced than $1/T$, where $T$ is the total duration of the data used.
The goal is to ensure the detection of a peak for a signal that is of
border-line statistical significance and to best locate the peak. The
frequency range we have searched is up to (or beyond) $\sim N/(2T)$,
where $N$ is the number of data points.

\subsection{Hard State}
 
\begin{figure}[f]
\begin{centering}
\epsfxsize=6.5in \epsfbox{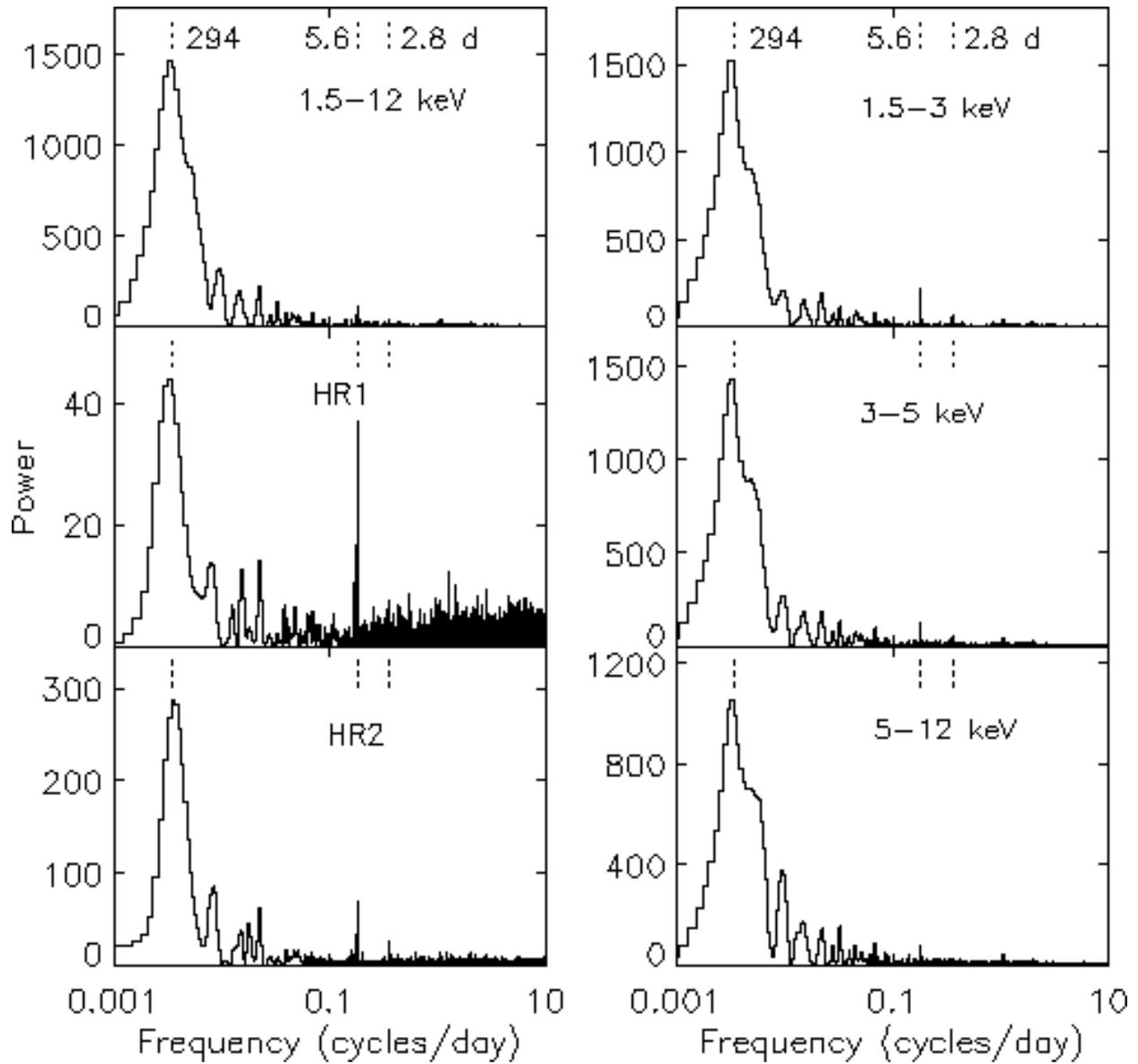}
\end{centering}
\figcaption[fig2.eps]{The Lomb-Scargle Periodograms  of the light curves and hardness ratios for the hard state. The periods of $2.8$ d, $5.6$ d, and $294$ d are marked by dotted lines. \label{power}}
\end{figure}

The Lomb-Scargle periodograms for the hard state are shown in Fig. \ref{power}. There is a distinct peak in the periodogram  at a frequency that is
consistent with Cyg X-1's   optically determined orbital period, i.e., $5.599829 \pm 0.000016$ days (\cite{brocksopp99}; see also \cite{gies82} and  \cite{lasala98}).  The peak is much more apparent in the periodograms of the hardness ratios than in those of the light curves.   Some of the periodograms also have a significant peak at the frequency of $1/2.8$ d$^{-1}$, the first harmonic of the orbital period.  There is a large peak at a very low frequency corresponding to a period of about $300$ days, which is consistent with the  reported $294 \pm 4$ day period by Priedhorsky, Terrell, \& Holt (1983) and by Kemp et al. (1983). However, the temporal span of the  ASM data is too short to  confirm this period; it could simply be ``red noise''.  In fact, this peak is no longer distinct in the periodograms calculated using an extended set of data, i.e., 860 days of hard state  data.  No other periodicities stand out at frequencies less than $20$ cycles per day except for the ``peaks'' at $ \lesssim 0.1$ cycles per day which appear to be  red noise (the spectrum for frequencies larger than $10$ cycles per day is not shown).

\begin{figure}[f]
\begin{centering}
\epsfxsize=6.5in \epsfbox{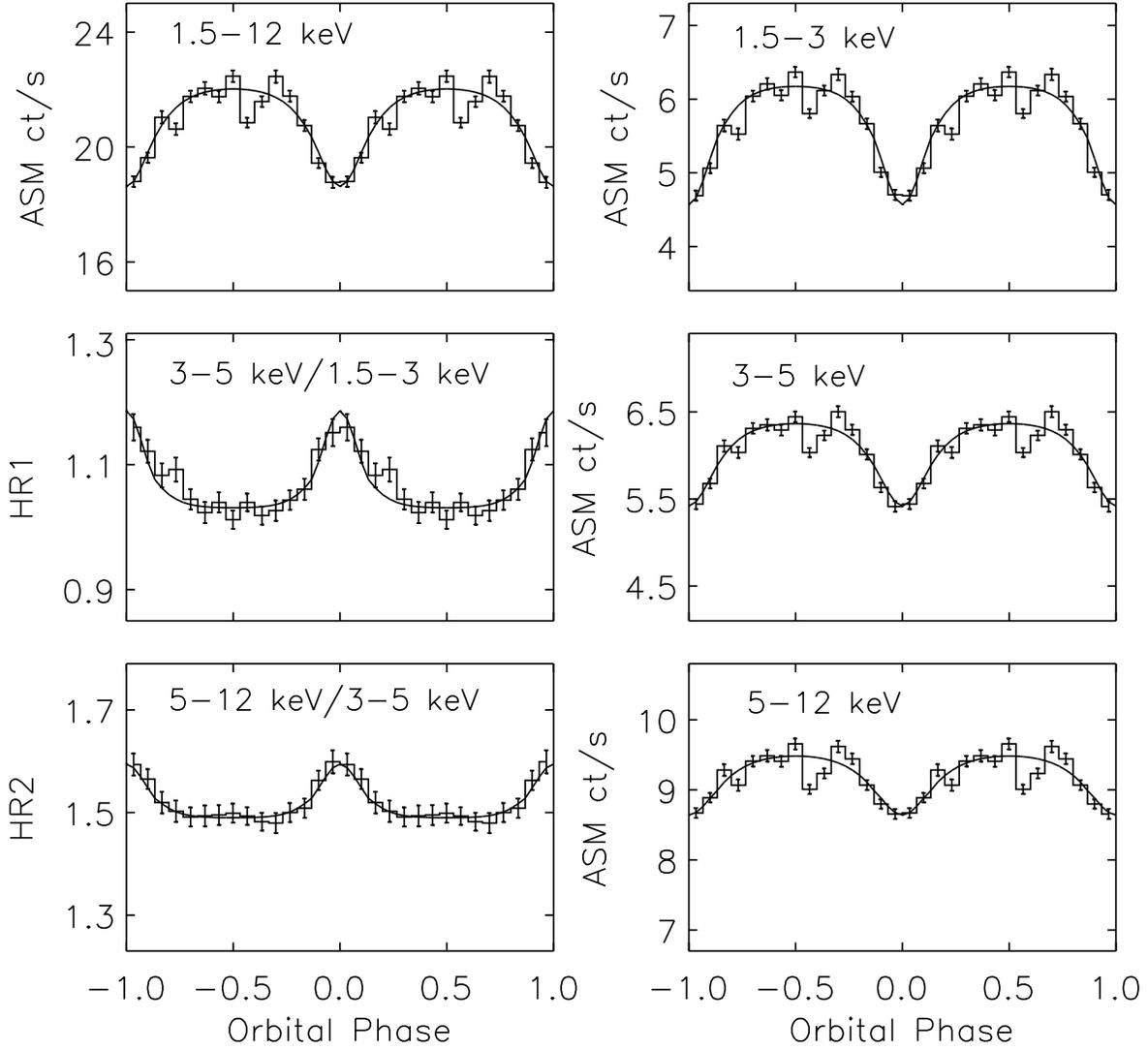}
\end{centering}
\figcaption[fig3.eps]{Folded light curves and hardness ratios for the hard
state.  The histograms represent the observations.  Orbital phase zero is defined as the superior conjunction of the X-ray source. The error bars represent one
standard deviation. The
smooth curves show the predictions from a wind absorption model for $i=30^\circ$ (see text).  \label{fold}}
\end{figure}

The data were folded modulo the orbital period of $5.599829$ days to study the phase-dependent variations (Fig. \ref{fold}). We used the orbital ephemeris  reported recently by Brocksopp et al.  (1999).    The most distinctive feature in the folded light curves is
the broad intensity dip.  It is seen in all energy bands and is  centered on the superior conjunction of the
X-ray source (phase zero).  The dip profiles are quite symmetric
about superior conjunction. The
fractional amplitude of the modulation in the light curves is larger in the lower energy
band, which manifests itself as gradual
spectral hardening during the dip.  The
fractional amplitudes of the dip relative to the average non-dip
intensities (phase $0.3$-$0.7$)
are $23$\% for $1.5$--$3$ keV, $14$\% for $3$--$5$ keV, and $8$\% for
$5$--$12$ keV.   The widths (FWHM) are all about 27\% of the orbital  period.  The corresponding fractional changes 
of HR$1$ and HR$2$ are about $13$\% and $8$\% respectively with
similar widths.   Taking
into account the variation  of the non-dip
intensity in the folded light curves, we estimated the 
uncertainty in the fractional orbital modulations to be less than
$4\%$.

\begin{figure}[f]
\begin{centering}
\epsfxsize=6.5in \epsfbox{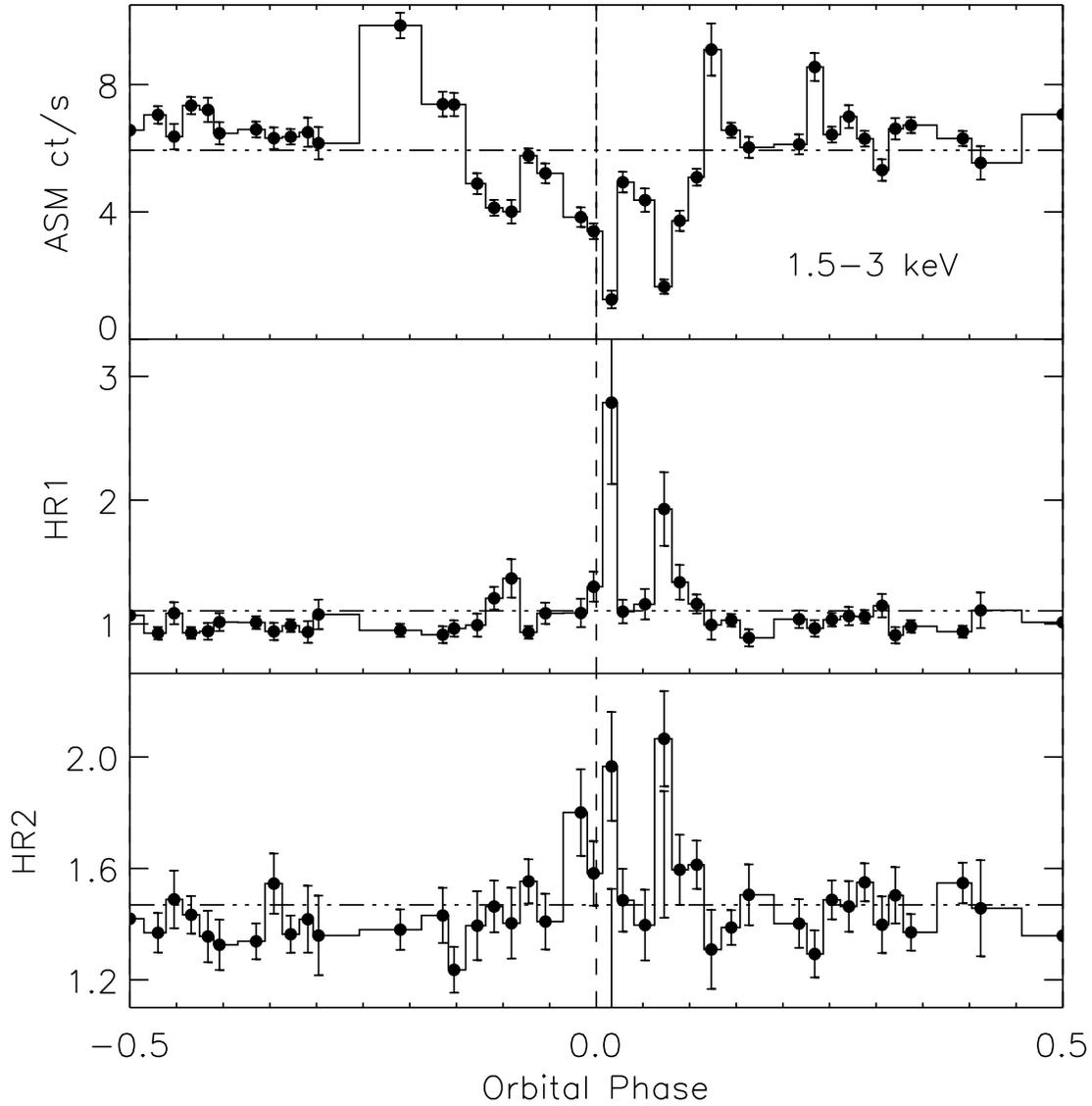}
\end{centering}
\figcaption[fig4.eps]{Light curve and hardness ratios for one
orbital cycle of the hard state.  The data are averaged in $0.1$-day bins. Some phase bins contain  no data points.  The histograms and the overall average (horizontal lines) are to aid the eye. Note the complex structures around phase $0$. \label{dip}}
\end{figure}

Complex structures are evident within the dip for at  least  $25\%$ of the orbital cycles observed by the ASM. The profile of the structure also seems to vary from cycle to cycle.  As the ASM data are unevenly  sampled in time, we have found only a few orbital cycles that are relatively uniformly sampled. We show in Fig. \ref{dip}  one such cycle  of the hard
state observations with time
bins of $0.1$ day in the energy band $1.5$--$3$ keV. There is a
broad intensity dip at superior conjunction with substantial
substructure.  In particular, there are two 
narrow  dips  near superior conjunction:  within a few hours, the intensities dropped by a factor of $\sim 2$, and the hardness ratios (HR1) increased by a factor of $> 1.6$.  This indicates that the dips were much less pronounced at higher energies as might be expected from an absorption process.
These smaller dip-like structures are similar to those reported
from previous missions.  It is possible that the broad dip may be, partially or wholly, due to
the superposition of smaller dips. We do not explore this
possibility further because the study of such small dips (at time scale of seconds to hours) requires more frequent sampling around superior conjunction than is  provided by the ASM.  

\subsection{Soft State}

\begin{figure}[f]
\begin{centering}
\epsfxsize=6.5in \epsfbox{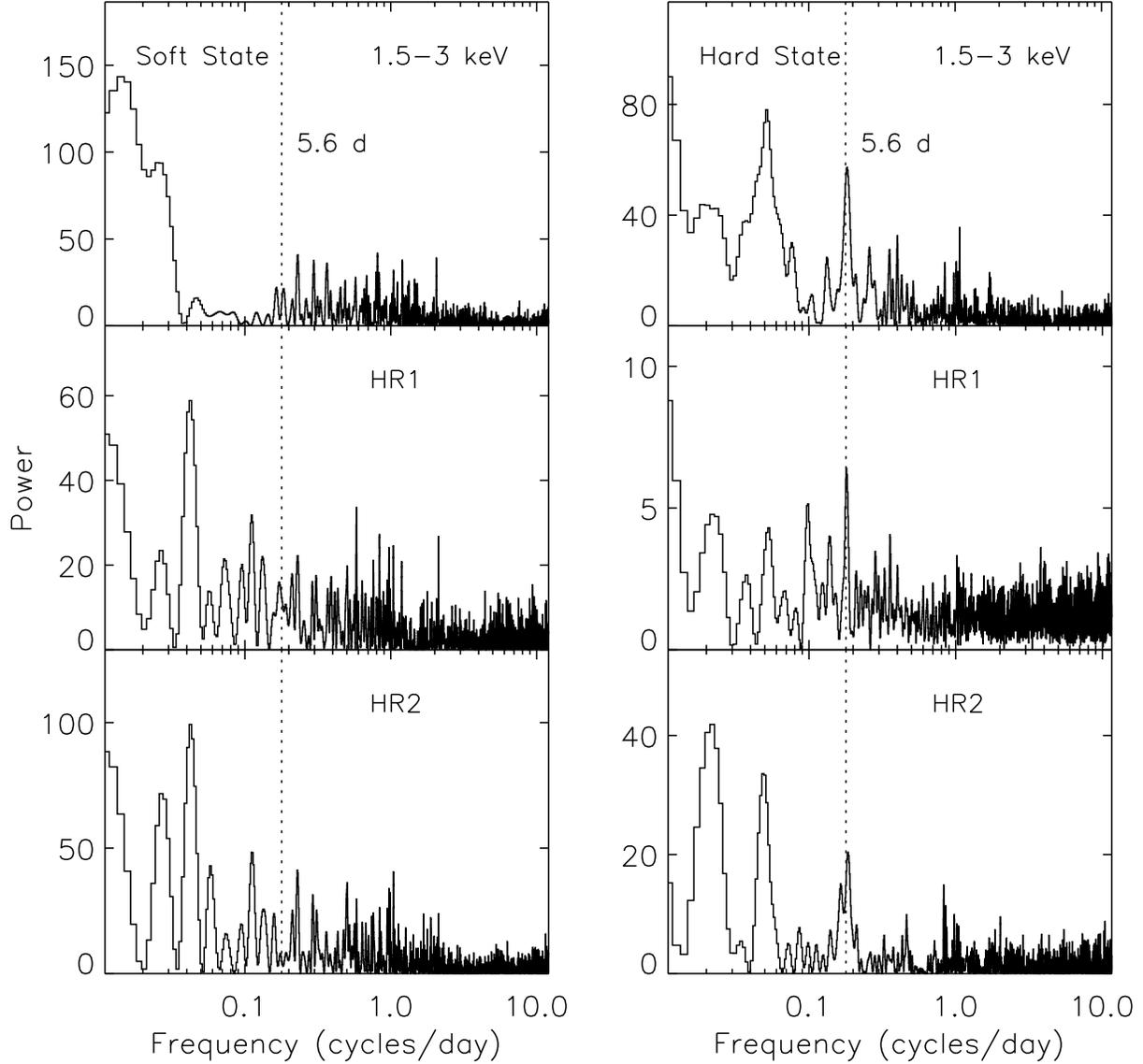}
\end{centering}
\figcaption[fig5.eps]{Lomb-Scargle periodograms for the soft (left) and hard state (right).  The time interval for the hard state was selected to contain a number of data points comparable to that for the soft-state time interval.  The $5.6$-day orbital period is detected in the hard-state data but is not apparent  in the soft-state data. \label{power_comp}}
\end{figure}

 In contrast with the hard state, Lomb-Scargle periodograms of the
soft-state data show no large power at the orbital period compared
with the neighboring powers (Fig. \ref{power_comp}, left
panel). Neither were any other periodicities found in the frequency
range of $0.1$--$10$ cycles per day.  At low frequencies,
i.e. $\lesssim 0.1$ cycles per day, red noise is evident. For a direct
comparison of the soft state data with the hard state data, we
constructed periodograms for an $80$-day segment of the hard-state
data that has a comparable number of data points
(Fig. \ref{power_comp}, right panel).  The $5.6$ day orbital period is
clearly detected in the hard state periodogram but is not obvious in the soft state.

\begin{figure}[f]
\begin{centering}
\epsfxsize=6.5in \epsfbox{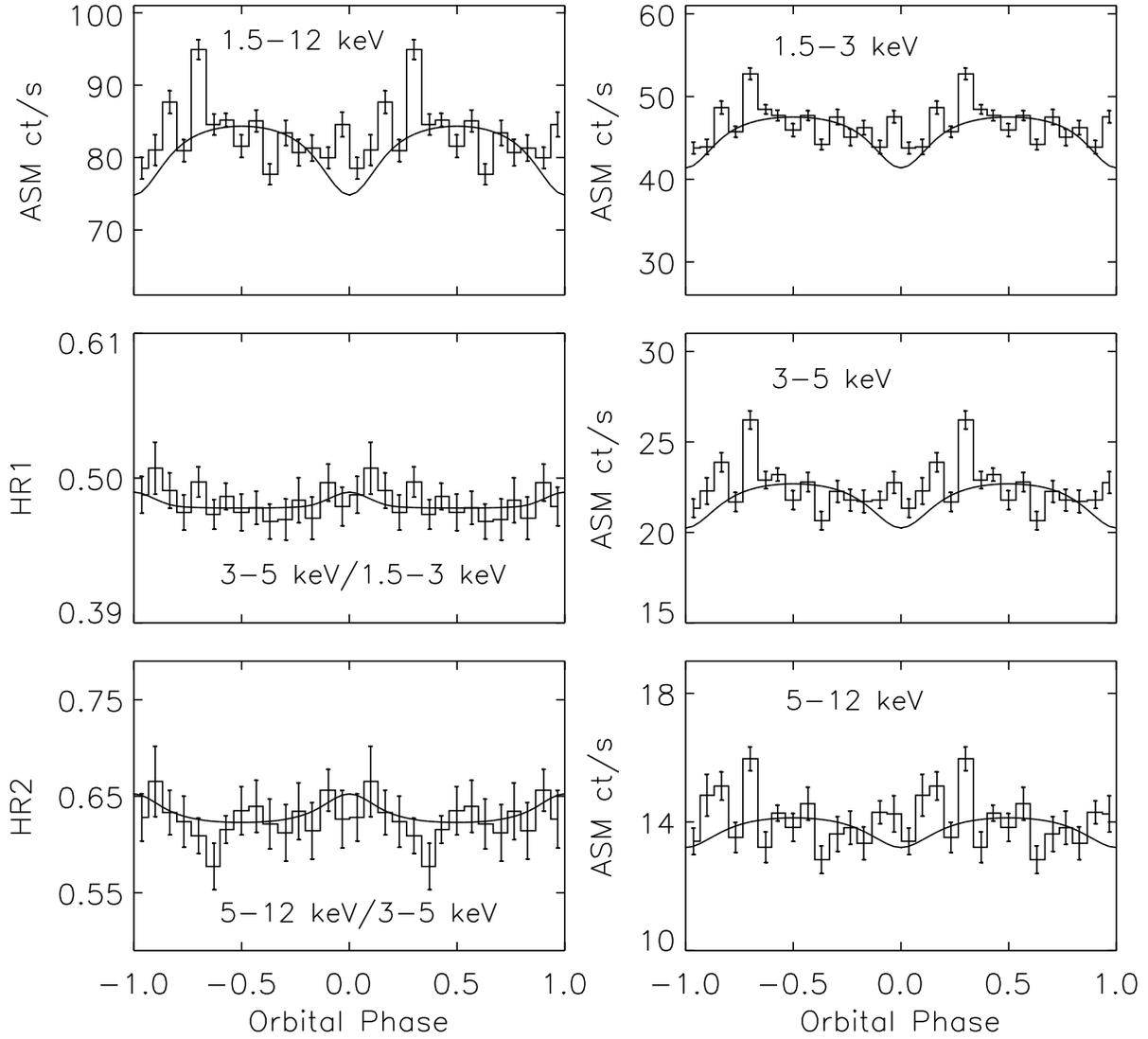}
\end{centering}
\figcaption[fig7.eps]{Folded light curves and hardness ratios, as in Fig. 3, but for the soft state.  The simple wind absorption  model produces lower fractional  orbital modulation in the soft state compared with the hard state. However, a reduction of the wind density is  required to explain the soft state data. \label{fold_high}} 
\end{figure}

In comparing periodograms, we use the normalized variance, i.e., the
observed total variance of the count rate divided by the average
rate.  For a sinusoidal modulation superposed on random noise, the
expected height of a peak in the periodogram is then proportional to
the product of the number of data points and square of the fractional modulation
divided by the normalized variance (see equation 21 in Horne \&
Baliunas 1986).  For the $1.5$--$3$ keV band soft state data, the
normalized variance is $\sim (0.53)^2 $ that of the hard-state data. Thus,
for comparable fractional orbital modulations (assumed to be nearly
sinusoidal), we expect the signal power of the soft state to be $\sim
(0.53)^{-2} \approx 3.6$ times that of the hard state,  or $P \sim 204$
(Fig. \ref{power_comp}).  The absence of a peak with $P > 25$ at the
orbital frequency in the soft state data thus clearly excludes the presence of
comparable fractional orbital modulations in the $1.5$--$3$ keV light
curves of the two states.  The folded orbital light curves and
hardness ratios for the soft state (Fig. \ref{fold_high}) also fail to
reveal any significant broad dip or spectral hardening near superior
conjunction.

A quantitative comparison of the fractional rms amplitude of the X-ray
orbital modulation of the two states was estimated from the classic periodogram (power density  spectrum estimated using FFT)  of the same data used above.  We have binned the data and
filled data gaps with the average rate in order to apply the FFT.  It is well known that the rms variation of the source signal in the data can be estimated using the FFT power spectrum assuming that the signal power can be properly separated out from the total power spectrum (cf., \cite{lewin88}; \cite{klis89}).  It is therefore relevant  to study the distribution of the noise power in the periodogram.  In the analysis below, only the powers at frequencies
$\ge 0.1$ cycles per day were considered because the noise power spectrum is relatively flat in this region.  The  powers were first
divided by the local mean which was obtained from a
linear fit to the power as a function of frequency. The scaled noise powers of both the soft and
hard state data were found to be consistent with a $\chi ^{2}$
distribution with $2$ degrees of freedom. We then assumed that modulation
at the orbital period would yield peaks with the same widths in the
power density spectrum from both states. On this basis, we derived the fractional rms amplitude of
the orbital modulation for the hard state and an  upper limit for the soft
state for each ASM energy band at more than $90\%$ confidence (cf., \cite{lewin88}; \cite{klis89}). This procedure
was repeated for different time-bin sizes ($0.0625$, $0.125$, $0.25$ and $0.5$ days) to check for consistency of
the results.  We found that in the $1.5$--$3$ keV band, the fractional
rms amplitude of the orbital modulation for the soft state is at most
$33\%$ of that for the hard state. The $3$--$5$ and $5$--$12$ keV bands yield higher percentages.



\section{Models}

The broad dip in the folded light curves  cannot be attributed to a partial eclipse by the companion.  The companion is a supergiant with a size more than $10^3$ times larger than the X-ray emitting region, 
so an eclipse of duration nearly $27\%$ of the orbital
period would have to be total.  Neither can the dip be caused by absorption by neutral material with solar elemental abundances since the observed $8\%$ reduction in  flux in the $5$--$12$ keV band  would then be accompanied by  a flux decrease in
the $1.5$--$3$ keV band of more than $80\%$  as opposed to the
observed $23\%$.

We have modeled the broad dip assuming that it is produced by absorption and scattering of the X-rays by a smooth isotropic stellar wind from
the companion star. The wind is partially ionized by the X-ray
irradiation.  The X-ray modulation is then caused by changes in the optical
depth along the line of sight to the black hole as a function of orbital phase. For simplicity, we did not consider possible complex structures in the wind,  e.g., the  tidal streams which could account for the strong X-ray attenuation  at late orbital phases ($> 0.6$) in some other wind accreting systems (e.g., \cite{blondin91}). In our calculation, we neglected the
influence of the UV emission from the optical star upon the ionization state of the wind as we expect it to have  little
effect on the X-ray opacity in the ASM energy band.

The radiatively driven wind model of Castor, Abbott, \& Klein
(1975) was adopted in our calculation. In this model, the velocity of the wind can be described by a simple power law for $R > R^*$:
\begin{equation}
\upsilon _{wind}=\upsilon _{\infty}[1-\frac{R^*}{R}]^ {\alpha},
\label{velocity}
\end{equation}
where $\upsilon _{\infty}$ is the terminal velocity of the wind, $R$ the distance from the center of the star, $R^{*}$ the
radius of the star,  and $\alpha$ a fixed index.  A spherically symmetric wind  is assumed for simplicity.  We therefore approximate the wind density profile as:
\begin{equation}
n(R) = [{\frac{R^{*}}{R}}]^2  \frac {n_0} { \{1-[R^{*}/R]\}^{\alpha} },
\label{density}
\end{equation}
where $n(R)$ is the number density of the wind, and $n_0$ is a wind density
parameter, expressed in terms of the proton number density.  The mass
loss rate by the wind thus is  $\dot{M}=m_{H}n_0 \times {4\pi {R^{*}}^2\upsilon
_{\infty }}$, where $m_{H}$ is the atomic hydrogen mass.

Simulated ASM light curves
in the energy band $E_1$--$E_2$ were produced by integrating along the
line of sight from the black hole for a given orbital phase $\phi$:
\begin{eqnarray}
I (\phi) =\int ^{E_2} _{E_1} dE I_{0} (E) Q(E) \underbrace{\exp{[-{\int ^{r_2} _{r_1}  
n( R(\phi,r,i))\times \sigma (E,\zeta)}dr]}}_{wind\  absorption}\underbrace{\exp{[-N_{H}\times \sigma _{0} (E)]}}_{interstellar \  absorption},
\label{I}
\end{eqnarray}
where $I_{0} (E)$ is the intrinsic X-ray energy spectrum, $Q(E)$ is
the ASM energy-dependent detection efficiency, $r$ is the distance
from the X-ray source, $\sigma (E,\zeta)$ is the photoelectric
absorption cross-section per hydrogen atom for the partially ionized
gas as a function of the energy and the ionization parameter $\zeta
=L_{x}/[nr^2]$ where $L_{x}$ is the effective source luminosity
between $13.6$ eV and $13.6$ keV, $\sigma _0(E)$ is the absorption
cross section per hydrogen atom for neutral gas, and $N_{H}$ is the
interstellar hydrogen column density.

The values for the parameters in equations (\ref{density}) and
(\ref{I}) were determined or adopted as follows.  For the wind model,
we took $\alpha = 1.05$, $R^{*} =1.387 \times 10^{12}$ cm, and
$\upsilon _{\infty}=1586$ km s$^{-1}$ from Gies \& Bolton (1986b) who
fitted equations (\ref{velocity}) and (\ref{density}) to the numerical
results from Friend \& Castor (1982) for the radiative-driven wind
profile of Cyg X-1. These values are for binary separation $a=2R^{*}$,
corresponding to a $98\%$ Roche lobe fill-out factor of the companion,
and for a wind profile that resembles a smooth wind from a single
O$9.7$ I supergiant.  The shape of $I_{0} (E)$ was chosen to be
similar to that seen in the ASCA observations of Cyg X-1 in the hard
state, i.e. with blackbody and broken power law components
(\cite{ebisawa96}).   The binary
inclination angle was taken as $i=30^{\circ}$ from the most probable
value derived by Gies \& Bolton (1986a). The interstellar hydrogen
column density was taken as $N_{H}=5\times10^{21}$ cm$^{-2}$, slightly
less than the values used in Ebisawa et al. (1996), {for a better fit
to the ASM data} .  The values for $\sigma _0(E)$ are from Morrison \&
McCammon (1983), and finally, for the wind, solar elemental abundances
were assumed (as listed in \cite{morrison83})( Table \ref{tab1}).


The cross-section $\sigma (E,\zeta)$ depends highly on the ionization
state of the wind.  Under the assumption of a steady state, the
ionization state of an optically thin gas illuminated by an X-ray
source can be uniquely parameterized by the ionization parameter
$\zeta$ for a given X-ray source spectrum (\cite{tarter69}). For each
ionization state of the optically thin gas, the local effective X-ray
opacity can be uniquely determined from atomic physics calculations.
The program XSTAR (v. 1.46,  see \cite{kallman82} for theoretical
basis) was used to obtain an opacity table which contains $\sigma
(E,\zeta)$ for a wide range of ionization parameters and energies.
For any particular ionization parameter value, $\sigma (E,\zeta)$ can
be constructed by interpolation. In our model,  the Thomson scattering
cross section was added to the cross section derived with the use of
XSTAR.  The wind absorption factor in equation (\ref{I}) was
integrated over the range $10^{11}$ cm $<r<10^{13}$ cm.  For
$r<10^{11}$ cm the wind is highly ionized while for $r>10^{13}$ cm the
density of the wind becomes very small; thus the absorption of the
X-rays by the wind is negligible in both cases.

Our procedure was to find the wind density parameter $n_0$ which
produced light curves with fractional orbital modulations matching
those obtained from the hard state ASM data.  The spectral parameters
from Ebisawa et al. (1996) were also adjusted slightly to match the
intensity levels in the three ASM energy bands.  The best-fit
values were obtained by minimizing the $\chi ^2$ values of  each model light curve relative to the data. The range of acceptable fit, with
$>90\%$ confidence level, is then estimated based on the increase of
$\chi ^2$ from the minimum (cf., \cite{lampton76}).  Note the
uncertainties in our results do not include the uncertain effects of
the assumptions and binary parameters adopted for the calculation. The
expected variance of the data in calculating $\chi ^2$  was taken to be  the variance of the
data between phase 0.3--0.7 to account for the possible intrinsic
uncertainty associated with the data.

The best-fit model light curves of the hard state for our choice of
$i=30^{\circ}$ are plotted as the solid lines in Fig.  \ref{fold} to
compare with the observational data.  Clearly this simple wind model
can account for the observed X-ray orbital modulation in the hard
state very well.  The adjusted spectral parameters are listed in Table
$1$.  For a distance of $2.5$ kpc, the derived intrinsic $1.3$--$200$
keV X-ray luminosity from this model is $6.6 \times 10^{37}$ ergs
s$^{-1}$, which is consistent with the previously reported value
(e.g., \cite{zhang97}).  The wind density parameter is estimated to be $n_0
= (6 \pm 1) \times 10^{10}$ cm$^{-3}$, 
indicating a total mass loss rate of $\sim 6\times 10^{-6}$\msun per
year. This value of $n_{0}$ is a factor of $\sim 3 $ larger than
that determined by Gies \& Bolton (1986b).  The hydrogen
(neutral+ionized) column density is about $1.0\times 10^{23}$
cm$^{-2}$. The ionization parameter $\zeta$ varies from $10^5$ to
$10^2$ along the integration path.  

\begin{figure}[f]
\begin{centering}
\epsfxsize=6.5in \epsfbox{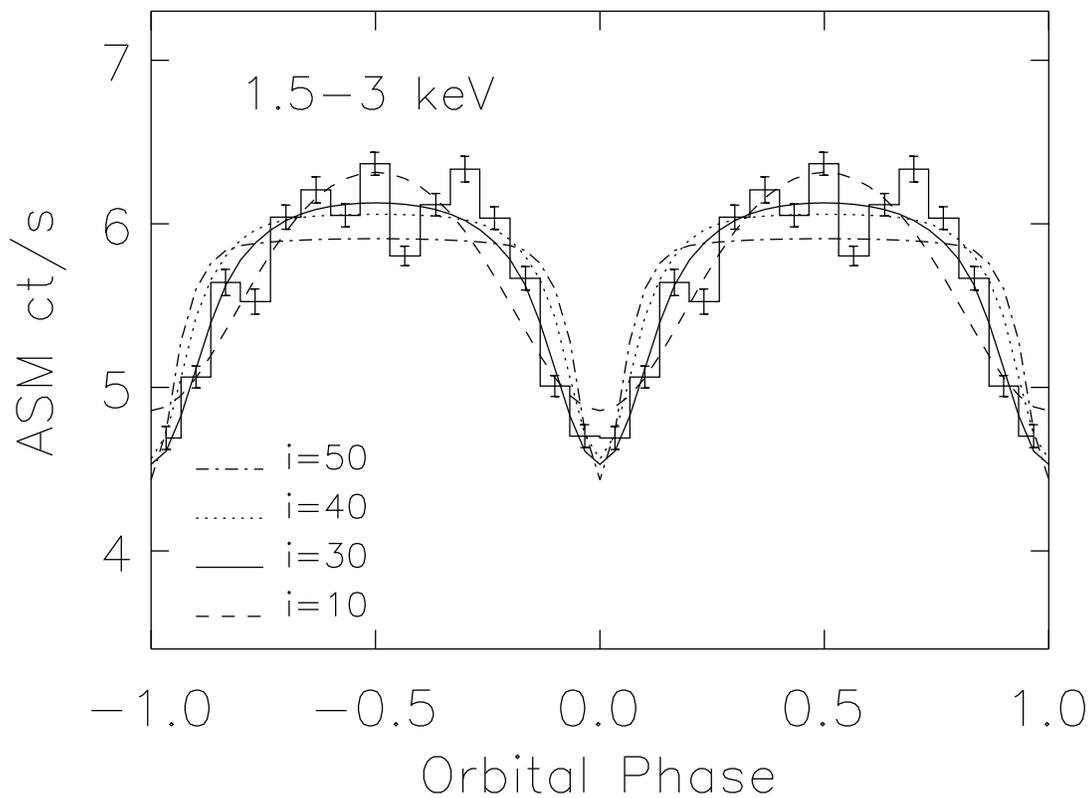}
\end{centering}
\figcaption[fig6.eps]{Predicted hard-state light curves from the wind absorption model for inclination angles between $10^\circ$--$50^\circ$.  The width of the dip decreases for larger inclination angles. \label{wind_comp} }
\end{figure}

We have studied the effect of the inclination angle $i$ upon the quality of the fit of the model light curves to the data.  The minimum of the $\chi ^2$ value
was found at roughly $i=30^\circ$ for data of all three energy bands.
The acceptable range of the inclination angle was found to be
$10^\circ \lesssim i \lesssim 40^\circ$, determined primarily by the
data in the 1.5--3 keV band.  This constraint is mainly due to the
fact that the width of the dip of a fixed fractional amplitude in our model
decreases if we increase the inclination angle (Fig. \ref{wind_comp}). The best-fit $n_0$
ranges from $3.7 \times 10^{10}$ cm$^{-3}$ for $i=40^\circ$ up to $1.6
\times 10^{11}$ cm$^{-3}$ for $i=10^\circ$.  Note that $n_0$ decreases
if we choose a larger inclination angle $i$.  The results are
relatively insensitive to the ASM efficiency $Q(E)$ and the intrinsic
X-ray spectral shape.

In the BATSE band, the X-ray opacity is entirely due to electron scattering. We found that in all our acceptable fits, attenuation 
caused by electron scattering would modulate the apparent intensity by 
$4$--$6\%$ peak-to-peak, which is in good agreement with the data (\cite{robinson96}). 

 We repeated the same procedure for the soft state for {\it
$i=30^\circ $}, assuming the same wind density profile (with
$n_{0}=6.0 \times 10^{10}$ cm$^{-3}$) and using the energy spectrum
adjusted slightly from that in Cui et al. (1997a), again to match the
count rates in the three energy bands.  The results are plotted as
solid lines in Fig. \ref{fold_high}.  The model produces light curves
of much smaller orbital modulations than in the hard state because the
wind is more ionized due to a much larger flux of soft X-ray photons
in the soft state.  In the 1.5---3 keV band, the amplitude of the
modulation ($\sim 14\%$) in the model light curve is not consistent
with the upper limit ($\sim 9\%$) determined in section 3.2.  Better
fits to the data can be found with smaller values of $n_{0}$.  For
$n_{0} < 4 \times 10^{10}$ cm$^{-3}$, the fractional modulation of the
model light curve is $< 9 \%$ in the $1.5$--$3$ keV band for the soft
state, which is consistent with the upper limit. A wind model with a
non-variable wind density therefore cannot explain the data in the
hard and soft states simultaneously in case of {\it $i=30^\circ $}.
 However, the non-detection
of the orbital modulation in the soft state can be explained if the wind density is
reduced by a factor of about 2 relative to the hard state.

\begin{figure}[f]
\begin{centering}
\epsfxsize=6.5in \epsfbox{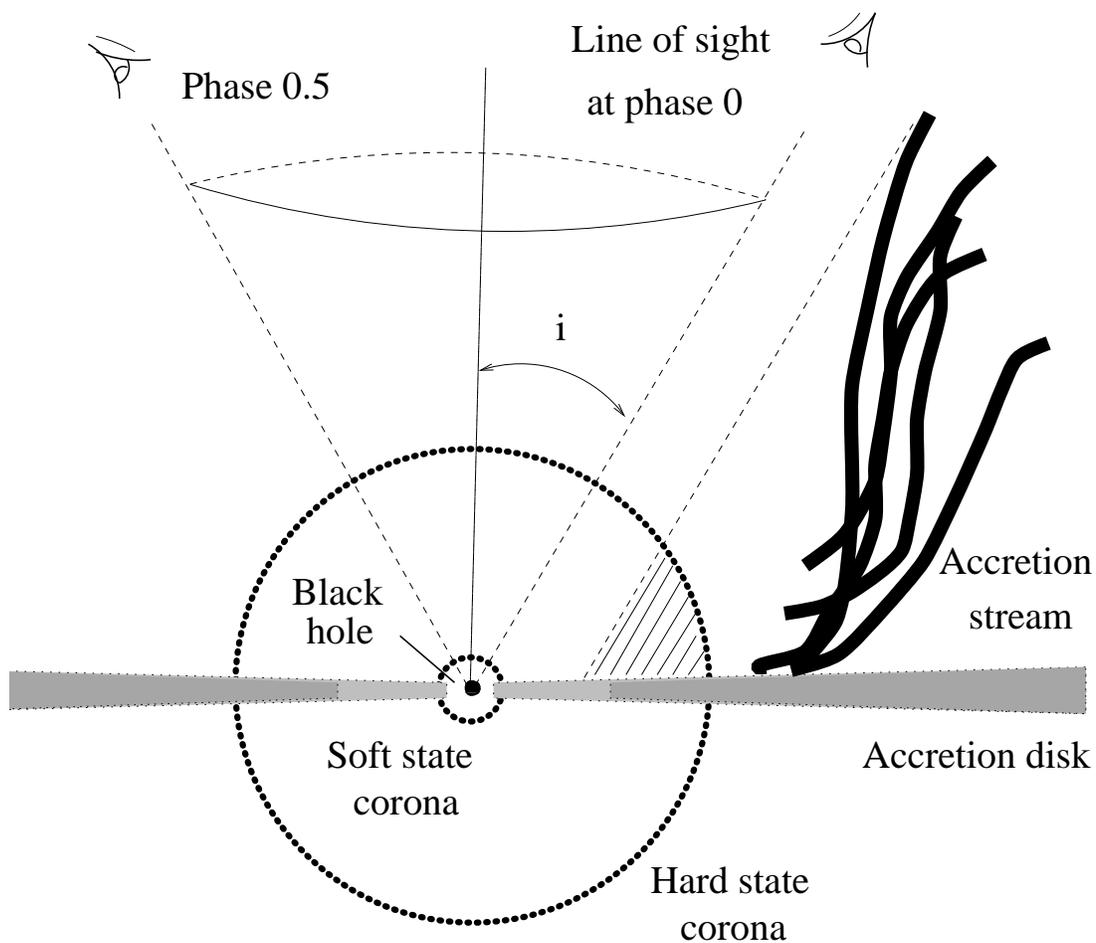}
\end{centering}
\figcaption[fig8.eps]{A schematic illustration of a possible geometry of the X-ray emitting region of Cyg X-1. Note that the size of the corona and the radius  of the inner edge of the disk are larger in the hard state than in the soft state. The observed X-ray orbital modulation in the hard state is attributed to the partial covering of a large corona (the shaded region). The lack of modulation in the soft state may due to significant shrinkage in the size of the corona. \label{corona}}
\end{figure}

Alternatively, the X-ray orbital modulation observed in the hard state
may be caused by partial covering of a central X-ray emitting region
by the accretion stream. The wind density in this model is
assumed to be much less than that required in the  model discussed above, 
therefore it does not contribute significant X-ray opacity.   Hard
X-rays are generally thought to be produced by upscattering of low
energy photons by electrons in a hot corona (e.g., \cite {nolan84}).
Observations seem to favor the geometry of a spherical corona centered
on the black hole plus a standard thin disk (Fig. \ref{corona})
(\cite{dove978}; \cite{gierlinski97};
\cite{poutanen97}).  Recent studies indicate that the size of the
corona in the hard state could be as large as $10^{9}$ cm
(\cite{hua98}) and that it may {\it shrink} by more than a factor of
$10$ as the soft state is approached (\cite{cui97b}; \cite{esin98}).
For Cyg X-1 in both states, the X-ray emission observed above $1$ keV
is primarily from the corona.  The accretion stream may have a scale
height above the disk such that, viewed along the line of sight near
superior conjunction of the X-ray source, it partially obscures the
outer region of the large corona in the hard state but does not do so
in the soft state because the corona is much smaller
(Fig. \ref{corona}). This constrains the distance of the absorber to
be a few coronal radii away from the black hole.  A covering factor
around $23\%$ is sufficient to explain the observed depth of the dip
in the hard state with a cold absorber of line-of-sight hydrogen
column density of ($1$--$3$) $\times 10^{23}$ cm$^{-2}$.  If we take the degree of ionization into account, the hydrogen column density
could be much higher, which may account for the observed modulation in
the BATSE band.
 
\section{Summary}  

Our analysis of {\it RXTE}/ASM observations of Cyg X-1 leads to the
following conclusions: There is a broad smooth dip in the folded
orbital light curves of Cyg X-1 in the hard state. The dip is
symmetric about superior conjunction of the X-ray source.  The depth of the dip
relative to the non-dip intensity is around $23\% $ in the $1.5$--$3$ keV band,
$14\%$ in the $3$--$5$ keV band, and $8\%$ in the $5$--$12$ keV band. The FWHM of the
dip is $27\%$ of the orbital period in the energy range $1.5$--$12$
keV. Individual light curves show complex structures around superior conjunction in the form of dips of shorter duration.  Finally, no evidence is
found for orbital modulation during the $1996$ soft state of Cyg X-1.

We examined the possibilities  that the broad dip is produced by the absorption of the X-rays
by a stellar
wind from the companion star. This model reproduces the
observed light curves of the hard state well for inclination angles $10^\circ \lesssim i \lesssim 40^\circ$ and can also explain the soft-state data  if there was a reduction in the stellar wind density for the duration of the soft state.   Alternatively, the observed X-ray modulation in the hard state  may be  mostly due to the partial obscuration of a central hard X-ray emitting region by the accretion stream. The lack of the observed  orbital
modulation in the soft state can be attributed to a  significant
shrinkage in the size of the X-ray emitting region such that it is no longer obscured by the accretion stream. This model requires the accretion stream to have specific geometric properties,  such as   its scale height, width, and orientation. In both models,  the required hydrogen column density can reproduce $\sim 5\%$ orbital modulation due to electron scattering as observed in the BATSE data (\cite{robinson96}). 

\acknowledgments 
We are very grateful to the entire $\it{RXTE}$ team at MIT for their support. We thank Saul Rappaport, Ron Remillard, and Shuangnan Zhang for many helpful discussions.  We also thank Tim Kallman and Patrick Wojdowski for their  help with using XSTAR program.

\pagebreak

\clearpage
\newpage

\begin{table}
\caption{ Spectral Parameters for the Hard State of Cyg X-1 \label{tab1} }
\begin{tabular}{cccccccc}
\tableline
\tableline
\multicolumn{2}{c}{Blackbody \tablenotemark{a}}&  \multicolumn{4}{c}{Broken Power Law \tablenotemark{b}} & Absorption \tablenotemark{c} & Luminosity\tablenotemark{d} \\
$N_{b}$  & $T_{e}$ (keV) &  $\alpha_{1}$  &  $E_{b}$ (keV)  &  $\alpha_{2}$ & $Np$  & $N_{H}$ & $L_{X}$ \\
\tableline
7.5 & 0.16 & 1.84 &   4  & 1.515 & 2.63 & 5.0 & 3.4 \\
\tableline
\end{tabular}
\tablenotetext{a}{ $N_{b}$: normalization in units of 10$^{36}$ ergs s$^{-1}$ at d=2.5 kpc, $T_{e}$: temperature }
\tablenotetext{b} { $E_{b}$: break energy,  $\alpha_{1}$: photon index below $E_{b}$, $\alpha_{2}$:  photon index above $E_{b}$, \\ $Np$:  normalization in units of photon s$^{-1}$ cm$^{-2}$ keV$^{-1}$ at 1 keV for $E \ge E_b$}
\tablenotetext{c}{ Interstellar hydrogen column density in units of 10$^{21}$ cm$^{-2}$}
\tablenotetext{d}{ Calculated in the 13.6 eV--13.6 keV band in units of $10^{37}$ ergs s$^{-1}$ }
\end{table}

\end{document}